\documentclass[12pt]{new_tlp}

\usepackage{latexsym}
\usepackage{graphicx}
\usepackage{amsmath, amssymb, theorem,graphics}
\usepackage[dvips]{color}
\usepackage{fancybox}
\usepackage{multirow}
\usepackage{enumitem}
\usepackage{bussproofs}

\newcommand{\ignore}[1]{}

\newtheorem{theorem}{Theorem}

\newtheorem{lemma}{Lemma}
\newtheorem{proposition}{Proposition}
\newtheorem{corollary}{Corollary}

\definecolor{grey}{rgb}{0.8,0.8,0.8}
\definecolor{lightgrey}{rgb}{0.92,0.92,0.92}

 \newcommand\ThanksRemy{ The
  research by R\'emy Haemmerl\'e has received funding from the
  Programme for Attracting Talent/young PHDs of the Montegancedo
  Campus of International Excellence \emph{PICD}, the European Union
  7th Framework Programme under grant agreement 318337 \emph{ENTRA},
  Spanish MINECO TIN'12-39391 \emph{StrongSoft} and TIN'08-05624
  \emph{DOVES} projects, and Madrid TIC-\-1465 \emph{PROMETIDOS-CM}
  project.}

\newcommand{\ba}{\begin{array}}
\newcommand{\ea}{\end{array}}
\newcommand{\bda}{\[\ba}
\newcommand{\eda}{\ea\]}

\newcommand{\prop}[0]{\Longrightarrow}
\newcommand{\simp}[0]{\Longleftrightarrow}
\newcommand{\vars}[0]{\mathit{vars}}
\newcommand{\set}[1]{\mathsf{#1}}
\newcommand{\tv}{\vars}

\newcommand\user{\mathit{usr}}

\date{}

\title{On Termination, Confluence and Consistent CHR-based Type Inference}

\author[Duck and Haemmerl{\'e} and Sulzmann]{
  GREGORY J. DUCK \\
  Department of Computer Science, National University of Singapore \\
  \email{gregory@comp.nus.edu.sg} \and R{\'E}MY HAEMMERL{\'E}\thanks{\ThanksRemy}
\\
  Universidad Polit{\'e}cnica de Madrid \& IMDEA Software Institute \\
  \email{Remy.Haemmerle@imdea.org} \and
  MARTIN SULZMANN \\
  Hochschule Karlsruhe - Technik und Wirtschaft
  \\ 
  \email{Martin.Sulzmann@hs-karlsruhe.de} }

\begin{document}

\bibliographystyle{acmtrans}
\maketitle

\begin{abstract}
We consider the application of Constraint Handling Rules (CHR) for
the specification of type inference systems, such as that used by
Haskell.
Confluence of CHR guarantees that the answer provided
by type inference is correct and consistent.
The standard method for establishing confluence relies on
an assumption that the CHR program is terminating.
However, many examples in practice give rise to non-terminating
CHR  programs, rendering this method inapplicable.
Despite no guarantee of termination or confluence, the
Glasgow Haskell Compiler (GHC) supports options that allow the user to proceed
with type inference anyway, e.g. via the
use of the \texttt{UndecidableInstances} flag.
In this paper we formally identify and verify a set of relaxed criteria,
namely \emph{range-restrictedness}, \emph{local confluence}, and
\emph{ground termination}, that
ensure the consistency of CHR-based type inference that maps to
potentially non-terminating CHR programs. \\

\noindent To appear in Theory and Practice of Logic Programming (TPLP).
\end{abstract}

\begin{keywords}
Constraint Handling Rules,
confluence,
termination,
type classes
\end{keywords}

\section{Introduction}

Constraint Handling Rules (CHR)~\cite{Frhwirth:2009:CHR:1618539} 
are a powerful rule-based programming
language for specification and implementation of constraint solvers.
CHR has many application domains, including
constraint solving~\cite{chr_survey},
type inference systems~\cite{DBLP:journals/jfp/SulzmannDJS07},
coinductive reasoning~\cite{Haemmerle11iclp},
theorem proving~\cite{duck12smchr} and
program verification~\cite{duck13heaps}. 
This paper concerns the application of CHR to \emph{type inference system}s
for high-level declarative programming languages such as
Haskell~\cite{Haskell98Book} and Mercury~\cite{mercury}.
In particular, type constraints imposed by 
\emph{type classes}~\cite{wadler-blott:ad-hoc} can 
be straightforwardly mapped into a set of CHR rules.
Type inference with type classes is then reduced to CHR solving.

For example, consider the following Haskell type class declarations
\begin{verbatim}
class Eq a where (==) :: a->a->Bool     instance Eq a => Eq [a] where ...
\end{verbatim}
The class declaration introduces some (overloaded) equality operator
\texttt{==} whose type is constrained by the type class
$(\mathtt{Eq~a})$.
The instance declaration states 
that we obtain equality among lists
assuming we supply equality on the element type.
Here the notation $\texttt{[}a\texttt{]}$ represents a list type
with element type $a$.
Following~\cite{DBLP:journals/jfp/SulzmannDJS07}, the above
maps to the following CHR simplification rule
\bda{rcl}
\texttt{Eq~[a]} & \simp & \texttt{Eq~a}
\eda

Type inference via CHR solving is performed
by (1) generating the appropriate constraints out of the program text,
(2) solving these constraints w.r.t.~the set of CHR rules derived
from class and instance declarations.
For example, consider the function \texttt{f} that tests if a list
\texttt{xs} is equal to a singleton list containing \texttt{y}.
\begin{verbatim}
f xs y = xs == [y]
\end{verbatim}
To infer the type of \texttt{f}
we (roughly) generate
$\texttt{Eq}~t_\mathit{xs}, t_\mathit{xs} = \texttt{[}t_y\texttt{]},
    t_f = (t_\mathit{xs} \rightarrow t_y \rightarrow \texttt{Bool})$
which reduces, via application of the above CHR rule and substitution, to
$\texttt{Eq}~t_\mathit{y}, t_\mathit{xs} = \texttt{[}t_y\texttt{]},
    t_f = (\texttt{[}t_y\texttt{]} \rightarrow t_y \rightarrow \mathit{Bool})$.
Hence, function \texttt{f} has type
$\forall a. \texttt{Eq}~a \Rightarrow \texttt{[}a\texttt{]} \rightarrow a \rightarrow \mathit{Bool}$.

This approach extends to richer sets of type class programs
such as multi-parameter type classes and functional dependencies~\cite{JonesESOP2000}.
The advantage is that important type inference properties such as decidability
and consistency can be verified by establishing the respective
properties for the resulting CHR rules.

The answer of type inference is guaranteed to be \emph{consistent}
if the set of CHR rules is confluent.
A terminating set of CHR  rules is \emph{confluent} if it reduces any given
goal to the same answer regardless of rule application ordering.
Earlier work~\cite{DBLP:journals/toplas/StuckeyS05,DBLP:journals/jfp/SulzmannDJS07}
identifies sufficient conditions on type class programs to guarantee
that the resulting CHR rules are confluent.
A critical assumption is that CHR rules are terminating.
In doing so, the proof of confluence can  be reduced to establishing
a weaker condition, namely \emph{local confluence}, via the application
of Newman's Lemma~\cite{newman}.

The problem is that the kinds of CHR programs that arise in practice often
violate the termination assumption. For example,
consider the following set of multi-parameter class and instance declarations
that incorporates a \emph{functional dependency}
$\texttt{a} \rightarrow \texttt{b}$
\begin{verbatim}
class F a b | a -> b   instance F Int Bool   instance F a b => F [a] [b]
\end{verbatim}
The functional dependency roughly states that: given a type-class constraint
$\mathtt{F}~a~b$, then the type $b$ is a functionally determined by $a$.
These class and instance declarations can be mapped to
the following CHR rules
\bda{c}
\mathtt{F}~\mathtt{Int}~b \simp b=\mathtt{Bool} ~~~~ ~~~~ ~~~~
\mathtt{F}~\texttt{[}a\texttt{]}~b \simp b=\texttt{[}c\texttt{]},~
    \mathtt{F}~a~c \\
\mathtt{F}~a~b, ~\mathtt{F}~a~c \prop b=c
\eda
The first two rules capture the two instances
and also enforce the functional dependency for the respective
instance.

At first glance, the above CHR program may appear to be terminating.
Indeed, any \emph{ground} constraint $(\mathtt{F~t_1~t_2})$ will be reduced
in a finite number of steps.
However, consider the \emph{non-ground} goal
$(\mathtt{F}~\texttt{[}a\texttt{]}~a)$,
where $a$ is some variable,
for which we find the following non-terminating derivation
\bda{l}
\mathtt{F}~\texttt{[}a\texttt{]}~a ~\rightarrowtail~
(\mathtt{F}~\texttt{[}b\texttt{]}~b,~a = \texttt{[}b\texttt{]})
    ~\rightarrowtail~
(\mathtt{F}~\texttt{[}c\texttt{]}~c,~a = \texttt{[}b\texttt{]},~
    b = \texttt{[}c\texttt{]}) ~\rightarrowtail~
\ldots
\eda
The above example represents a typical (albeit much simplified) example
that is found when applying type classes for expressive 
type reasoning~\cite{Hallgren00funwith}.
For example, consider the following more realistic example,
which encodes addition at the level of types
\begin{verbatim}
data Zero             data Succ n             class Add a b c | a b -> c   
instance Add Zero b b       instance Add a b c => Add (Succ a) b (Succ c)
\end{verbatim}
Here the non-ground goal $\mathtt{Add}~(\mathtt{Succ}~a)~b~a$ also
exhibits the same non-terminating behavior.

Fortunately, realistic programs will usually \emph{not}
yield devious non-terminating constraints such as
$(\mathtt{F}~\texttt{[}a\texttt{]}~a)$.
Hence, practical implementations of type inference systems, such as the
Glasgow Haskell Compiler~\cite{ghc},
typically enable the
user\footnote{Via GHC's \texttt{UndecidableInstances} flag.}
to proceed with type inference even though the corresponding CHR program
is potentially non-terminating.
If the flag is enabled, the type inference engine must compute
the answer within a fixed number of reduction steps, otherwise an
error is reported.

This paper is concerned with the correctness of the above
``practical implementations''.
Current CHR theory concerning confluence and consistency of CHR programs
explicitly assumes the program is \emph{terminating for all goals}, which is
simply not applicable under our setting.
Our main contributions are:
\begin{itemize}[noitemsep,topsep=0pt,parsep=0pt,partopsep=0pt]
 \item[-] We establish that range-restricted, ground confluent
    CHR programs are \emph{consistent} (Section~\ref{sec:consistency}).
    This extends the classical CHR consistency result
    from~\cite{Abdennadher99confluence}.
 \item[-]
    We establish that terminating goals are confluent for
    range-restricted, ground-terminating and
    locally confluent programs
    (Section~\ref{sec:confluence-terminating-goals}).
 \item[-] We discuss how these results apply to the GHC/type class setting
   (Section~\ref{sec:haskell}).
   In particular, we show that if type inference finitely terminates with
   an answer, then that answer is \emph{unique}.     
\end{itemize}
Section~\ref{sec:chr-type-classes} reviews background
material on CHR.
Section~\ref{sec:conc} summarizes related work and concludes.

\section{Constraint Handling Rules} \label{sec:chr-type-classes}

Throughout this paper we use \emph{Haskell type notation} to represent terms,
constraints and predicates.
Under this scheme:
\begin{itemize}[noitemsep,topsep=0pt,parsep=0pt,partopsep=0pt]
\item[-] functors (a.k.a. atoms) begin with an upper case letter
  (the opposite to Prolog);
\item[-] variables begin with a lower-case letter
  (the opposite to Prolog); 
\item[-] term arguments are separated by whitespace; and
\item[-] the special functor $\texttt{[a]}$ is shorthand for
  $(\mathtt{List~a})$ and $\mathtt{a}\texttt{=}\mathtt{b}$ for equality.
\end{itemize}
For example, the term $\texttt{p(X, q(Y), list(Z))}$ under Prolog syntax
would be represented as $(\texttt{P~x~(Q~y)~[z]})$ under Haskell type syntax.

Constraint Handling Rules (CHR) is a rule-based constraint rewriting
programming language designed for implementing constraint solvers.

We assume, as given, the following disjoint infinite sets of variables:
$\set{ProgVars}$, $\set{GlobalVars}$ and $\set{LocalVars}$.
We define:
\begin{align*}
    \text{Terms} & ~~~~
        t ::= v ~|~ F~t~\ldots~t \\
    \text{Built-in Constraints} & ~~~~
        b ::= \texttt{True}~|~\texttt{False}~|~t = t \\
    \text{User Constraints} & ~~~~
        u ::= C~t~\ldots~t \\
    \text{Constraints} & ~~~~
        c ::= b ~|~ u
\end{align*}
where $v$ is a variable from some variable set.
A \emph{substitution}  is a mapping from variables to terms.
We use the notation $\theta.X$ to represent a substitution $\theta$ applied to a
term or constraint $X$.
We respectively define $\set{Cons}(V)$ and $\set{Usr}(V)$ as the
set of all constraints and the set of all user-constraints over
the set of variables $V$.
There are two main types of CHR rules:
\begin{align*}
H ~\simp~ B ~~~~ (\text{Simplification})
    ~~~~ ~~~~ \text{and} ~~~~ ~~~~
H ~\prop~ B ~~~~ (\text{Propagation})
\end{align*}
where $H \in \mathcal{M}(\set{Usr}(\set{ProgVars}))$ and $B \in
\mathcal{M}(\set{Cons}(\set{ProgVars}))$\footnote{$\mathcal{M}(X)$ is
  the set of all multisets built form the set $X$.}. %
The local variables of a rule are those variables that appear in the
body but not in the head.  Logically simplification rules (resp.\
propagation rules) are understood as equivalence (implication) between
the head and the body where local variables are implicitly
existentially quantified.

Let $\set{StateVars} = \set{GlobalVars} \cup \set{LocalVars}$ and let
$S_1, S_2 \in \mathcal{M}(\set{Cons}(\set{StateVars}))$ then we
define $S_1 \equiv S_2$ as the least relation satisfying
\[
\mathcal{CT} \models (\exists_\set{LocalVars} : \user(S_1) \mathop{=} \user(S_2)
\wedge S_1) \leftrightarrow (\exists_\set{LocalVars} : \user(S_1) \mathop{=}
\user(S_2) \wedge  S_2)
\]
where $\mathcal{CT}$ is the theory of term equality %
and $\user(S)$ the user constraints of the state $S$.
The set of all \emph{CHR states} $\Sigma$ is defined as the quotient set
$\Sigma = (\mathcal{M}(\set{Cons}(\set{StateVars})) / \equiv)$.
Note that we usually represent a state $[S] \in \Sigma$ by an $S$, and
we often drop the braces $\{\ldots\}$ around sets of constraints,
i.e. we write $\mathtt{P}~a, a = \texttt{[}b\texttt{]}$ instead of
$\{\mathtt{P}~a, a = \texttt{[}b\texttt{]}\}$.
We will say that a rule is {\em purely built-in} if its body does not
contain any user constraints.

Operationally, CHR is the \emph{abstract rewriting system}
$\langle \Sigma, \rightarrowtail \rangle$, where 
binary relation $(\rightarrowtail) \in \Sigma \times \Sigma$ is the
CHR \emph{derivation step} defined as the least relation satisfying:

\begin{prooftree}
\AxiomC{$(H \simp B) \qquad
    \mathcal{CT} \models S \rightarrow (\theta.H = C) \qquad
    \mathcal{CT} \models \exists : C \land S$}
\UnaryInfC{$C \uplus S \rightarrowtail \theta.B \uplus S$}
\end{prooftree}
\begin{prooftree}
\AxiomC{$(H \prop B) \qquad
    \mathcal{CT} \models S \rightarrow (\theta.H = C) \qquad
    \mathcal{CT} \models \exists : C \land S \qquad
    C \uplus S \neq  \theta.B \uplus C \uplus S$}
\UnaryInfC{$C \uplus S \rightarrowtail \theta.B \uplus C \uplus S$}
\end{prooftree}
where $(\uplus)$ is multi-set union,
$\theta: \set{ProgVars} \rightarrow \set{StateVars}$ is a 
substitution mapping $\vars(H)$ to $\vars(C)$ and
$\vars(B) - \vars(H)$ to fresh variables from $\set{LocalVars} - \vars(C, S)$.

Note that there are many different definitions for the operational
semantics of CHR.
Our version does not keep propagation histories, and as such will only
terminate for propagation rules with built-in only bodies.
This is sufficient for our purposes.

Let $(\rightarrowtail^=)$ the reflexive closure of
$(\rightarrowtail)$, and let
 $(\rightarrowtail^*)$ be the transitive closure of $(\rightarrowtail^=)$.
A pair of states $S_1, S_2 \in \Sigma$ is \emph{join-able} if there exists an
$S' \in \Sigma$ and derivations $S_1 \rightarrowtail^* S'$,
$S_2 \rightarrowtail^* S'$.
An abstract rewriting system $\langle \Sigma, \rightarrowtail \rangle$ is:
\begin{itemize}[noitemsep,topsep=0pt,parsep=0pt,partopsep=0pt]
\item[-] \emph{terminating} if there is no infinite derivation
    $(S \rightarrowtail \ldots)$ for all $S \in \Sigma$.
\item[-] \emph{locally confluent} 
    if for all $S, S_1, S_2 \in \Sigma: S_1 \leftarrowtail S \rightarrowtail S_2$ then
    $S_1$ and $S_2$ are join-able.
\item[-] \emph{confluent}
    if for all $S, S_1, S_2 \in \Sigma:
    S_1 \leftarrowtail^* S \rightarrowtail^* S_2$ then
    $S_1$ and $S_2$ are join-able.
\end{itemize}
If a program $P$ is both locally confluent and terminating, then $P$ is
confluent~\cite{newman}.
Confluence implies \emph{logical consistency} of
$P$~\cite{Abdennadher99confluence,HLH11ppdp}.
That is, the logical reading of $P$ does not imply false.

Let $\mathcal{I}$ be any property over states such that:
for all $S, S' \in \Sigma$ where $S \rightarrowtail S'$, if
$\mathcal{I}(S)$ holds then $\mathcal{I}(S')$ also holds.
Then $\mathcal{I}$ is an \emph{observable invariant}~\cite{duck07obs}.
If we define $\Sigma_\mathcal{I} = \{S~|~S \in \Sigma \land \mathcal{I}(S)\}$
then program $P$ is respectively \emph{$\mathcal{I}$-terminating},
\emph{$\mathcal{I}$-locally-confluent}, and \emph{$\mathcal{I}$-confluent}
if the abstract rewriting system
$\langle \Sigma_\mathcal{I}, \rightarrowtail \rangle$
is terminating, locally-confluent and confluent.

We define the set of all
\emph{ground states} $\Sigma_g$ as the canonical surjection of
$\mathcal{M}(\set{Cons}(\emptyset))$ onto $\Sigma$.
A CHR program $P$ is \emph{range restricted} iff groundness is an
observable invariant.\footnote{
Note that our definition of \emph{range-restricted}-ness is more general
than the standard definition, i.e. that $\vars(B) \subseteq \vars(H)$ 
for all rules $(H \simp B)$ or $(H \prop B)$.}

\medskip

Before continuing we state the monotonicity of CHR transitions as
a number of proofs of the present paper rely on it.  This property
means that if a transition step is possible in a state, then it is
possible in any state that contains additional constraints.
\begin{proposition}[Monotonicity]
  Let $S$, $T$, and $U$ be three states such that $\vars(S,T) \cap
  \vars(U) \subset \set{GlobalVars}$.  If $S \rightarrowtail T$ holds,
  then so does $(S \uplus U) \rightarrowtail^= (T \uplus U)$.
 \end{proposition}

\section{Consistency of Ground Confluent CHR} \label{sec:consistency}

In our first result we show that ground-confluence with
range-restrictedness guarantees the logical consistency of programs.

\begin{theorem}[Consistency]
\label{theorem:consistency}
If  $P$ is range-restricted and ground confluent program, then it is
consistent.
\end{theorem}
\vspace{-1em}
\begin{proof}
Define
\(  \mathcal{H} =  \{c ~|~ (\{c\} \uplus S) \in \Sigma_g \text{ and } S
\rightarrowtail^* \mathtt{True} \} \).
To establish consistency of $P$, it is sufficient to show
$\mathcal{H}$ is an Herbrand model for both the constraint theory and
the logical reading of the program:

For the constraint theory, clearly $\mathtt{True} \in \mathcal{H}$ whilst
$\mathtt{False} \not\in \mathcal{H}$.  Now consider an
  equality constraint $t=s$ between two ground terms.  If $t$ is
  syntactically equal to $s$, then the derivation
  $(t=s) \rightarrowtail^* \mathtt{True}$ trivially holds,
  i.e\ $\mathcal{H} \models t=s$.
  Otherwise if $t$ syntactically differs from $s$, for any $S \in \Sigma$
  we have that
  $(\{t=s\} \uplus S) = \mathtt{False} \not \rightarrowtail^*\mathtt{True}$,
  i.e.\ $\mathcal{H} \not\models t=s$.

For the logic reading of a simplification rule $(H \simp B)$, we are
  required
  to show that $\mathcal{H} \models \forall (H \leftrightarrow
  \exists_{\vars(H)} B)$, or equivalently $\theta. H \subseteq I$ iff
  there exists $\rho$ that coincides with $\theta$ on $\vars(H)$ such
  that $\rho.B \subseteq \mathcal{H}$.  If $\theta.H \subseteq
  \mathcal{H}$ then for any $h \in H$ there exists $(\{\theta.h\}
  \uplus S_h)\rightarrowtail^* \mathtt{True}$ for some $S_h \in
  \Sigma_g$. Therefore by monotonicity for $S = \biguplus_{h \in H}\{
  S_h \}$, $(\theta.H \uplus S) \rightarrowtail^* \mathtt{True}$.  Let
  $S' \in \Sigma_g$ be the state obtained by applying the
  simplification rule on $(\theta.H \uplus S)$. By definition of
  $\rightarrowtail$, $\rho.B \subseteq S'$ for some $\rho$ that
  coincides with $\theta$ on $\vars(H)$.  Then $S' \rightarrowtail^*
  \mathtt{True}$ by ground-confluence, and therefore $\rho.B \subseteq
  \mathcal{H}$, i.e.  $\mathcal{H} \models \theta.B$.
Conversely, if $\mathcal{H} \models \rho.B$ then for any $b\in B$
there exists some $S_b \in \Sigma_g$ such that $(\{\rho.b\} \uplus S_b)
\rightarrowtail^* \mathtt{True}$.  Define $S = \biguplus_{b \in B}\{ S_b \}$,
then by monotonicity $(\rho.H \uplus S) \rightarrowtail^*
\mathtt{True}$, therefore $\rho.H \subseteq S \subseteq \mathcal{H}$,
i.e. $\mathcal{H} \models \theta.H$.

For the logical reading of a propagation rule ($H \prop B$) one
may consider the simplification $(H \simp H, B)$ and apply previous
case.
\end{proof}

Our consistency result is similar to the original result
from~\cite{Abdennadher99confluence}, except that it (1) uses a more
general definition of range restrictedness, and (2) assumes ground
confluence versus confluence.  Also note that ground confluence
follows from ground termination and local confluence, using Newman's
Lemma~\cite{newman}.

\section{Confluence for Terminating Goals}
\label{sec:confluence-terminating-goals}

A non-confluent, non-terminating CHR program $P$  
may still be confluent and terminating for
\emph{specific goals}.
For example, the CHR program from 
the introduction has both
non-terminating and terminating goals:
\begin{align*}
\mathtt{F}~\texttt{[}a\texttt{]}~a \mathop{\rightarrowtail}
\mathtt{F}~\texttt{[}b\texttt{]}~b, a \mathop{=} \texttt{[}b\texttt{]}
    \mathop{\rightarrowtail} \ldots ~ (\text{non-termination})
    ~~~~ ~
\mathtt{F}~\texttt{[}a\texttt{]}~\texttt{[}a\texttt{]} \mathop{\rightarrowtail}
    \mathtt{F}~a~a ~ (\text{termination})
\end{align*}
Since type inference is the same as CHR solving, the first goal is clearly
problematic.
On the other hand, the second goal always terminates, 
and thus is acceptable.

In the following, we identify sufficient conditions
which guarantee that if for goal $S$
we find \emph{some} derivation $S \rightarrowtail^* S'$
where $S'$ is a non-$\mathtt{False}$ final state
then (a) \emph{all} derivations starting from $S$ will terminate
and (b) these derivations lead to the same state $S'$.
Part (b) follows rather easily once we have (a).
Hence, we first  consider part (a).

\subsection{Universal Termination follows from Existential Termination}
\label{section:universal}

We distinguish between different types of termination.
\emph{Universal termination} means that all derivations from a state $S$ will
terminate, i.e. there does not exist an infinite derivation
$(S \rightarrowtail^* \ldots)$.
In contrast, \emph{existential termination} means that there exists a
terminating derivation from $S$, i.e. there exists at least one
derivation of the form $S \rightarrowtail^* T \not\rightarrowtail$.
For example, the goal
$(\mathtt{F}~\texttt{[}a\texttt{]}~\texttt{[}a\texttt{]})$
is both existentially and universally terminating.

Our main result is as follows:
Given a range-restricted, ground-terminating and locally-confluent
program $P$, then a given state $S$ is universally terminating if
it is existentially terminating to a non-\texttt{False} final state.

\begin{theorem}[Universal Termination]\label{th:universal}
  Let $P$ be a range-restricted, ground-term\-inat\-ing, and locally-confluent
  program.
  If $S$ is existentially terminating to
  a non-$\mathtt{False}$ final state, i.e.
  $S \rightarrowtail^* T \neq \mathtt{False}$ and $T \not\rightarrowtail$,
  then $S$ is universally terminating.
\end{theorem}
The proof of the Theorem \ref{th:universal} relies on the following
lemmas.
\begin{lemma}\label{lemma:subs}
If $S \rightarrowtail^* T$, $T \neq \mathtt{False}$ and
$T \not\rightarrowtail$, then there exists a ground substitution
$\theta$ such that $\theta.S \rightarrowtail^* \theta.T$,
$\theta.T \neq \mathtt{False}$ and
$\theta.T \not\rightarrowtail$.
\end{lemma}
\vspace{-1em}
\begin{proof*}
Let $\psi$ be the m.g.u.\ of the equations in $T$.
Let $\rho = \{x \mapsto c_x ~|~ x \in \set{Vars}\}$ be a
ground substitution mapping variables to fresh constants $c_x$.
Then define $\theta = \{x \mapsto \rho.\psi(x) ~|~ x \in \set{Vars}\}$
and we see that:
\begin{itemize}[noitemsep,topsep=0pt,parsep=0pt,partopsep=0pt]
\item[-] $\theta.S \rightarrowtail^* \theta.T$ by monotonicity;
\item[-] $\theta.T \neq \mathtt{False}$ since $\theta$ is a unifier of
    the equations in $T$; and
\item[-] $\theta.T \not\rightarrowtail$ otherwise $T \rightarrowtail$ since
    $c_x$ were fresh constants. \quad $\mathproofbox$
\end{itemize}
\end{proof*}
\begin{lemma}\label{lemma:propagations_terminates}
  A set $P$ of purely built-in propagation rules  is
  terminating.
\end{lemma}
\vspace{-1em}
\begin{proof}\newcommand\rank{\mathit{rank}}
  Let $S$ be a state. Now consider all pairs $((H \prop B), \psi)$ such
  that $(H \prop B)$ is a rule of $P$ and $\psi$ is the m.g.u.\
  between $H$ and some subset $S'$ of $S$ (i.e.\ $S' \subseteq S$ and
  $\psi.H = \psi.S'$). There exists at most finitely many such pairs
  which we may enumerate thusly:
 \[
 ((H_1 \prop B_1), \psi_1), \dots, ((H_n \prop B_n), \psi_n)
 \]
 where the local variables of the $(H_i \prop B_i)$ have been
 previously renamed apart.  %
 Now let define the ranking of $S$ as $\rank(S) = n - |\{ i \mid
 \mathcal {CT} \models S \rightarrow \psi_i.B_i \}|$ where $|X|$ is
 the cardinality of the set $X$.  %
 One verifies that if $S \rightarrowtail T$ then $\rank(S) > \rank(T)$.
 It follows that $\rightarrowtail$ is terminating.
\end{proof} 
\begin{lemma}\label{lemma:inf}
  Let $P$ be a range-restricted and ground-terminating program. %
  Suppose there exists an infinite derivation $(S \rightarrowtail^*
  \ldots)$ with $\vars(S) \subseteq \set{GlobalVars}$.  Then for all
  ground substitution $\theta$ with domain $\set{GlobalVars}$, we have
  that $\theta.S \rightarrowtail^* \mathtt{False}$.
\end{lemma}
\vspace{-1em}
\begin{proof}
Assume there exists an infinite derivation of the form: 
\[
S  \rightarrowtail S_1  \rightarrowtail \cdots \rightarrowtail S_n \rightarrowtail \cdots
\]
Let $\rho$ be an arbitrary ground substitution with domain
$\set{GlobalVars}$.  By monotonicity:
\[
\rho.S  \rightarrowtail^=
\rho.S_1 \rightarrowtail^=
\cdots \rightarrowtail^=
\rho.S_n  \rightarrowtail^= \cdots
\]
Since $P$ is ground terminating, we get that there exists some $i \in
\mathbb N$ such that for any $j \geq i$, $\rho.S_i = \rho.S_j \not
\rightarrowtail$.  %
By Lemma \ref{lemma:propagations_terminates}, there is a $k \geq i$
such that the transition step $S_k \rightarrowtail S_{k+1}$ is induced
by a simplification rule or a propagation rule with user-defined
constraints in the body. %
In such a case, we observe if $S_k \rightarrowtail S_{k+1}$ and
$\rho.S_k \not \rightarrowtail \rho.S_{k+1}$ then $\rho.S_k =
\mathtt{False}$.
\end{proof}  
\vspace{-1em}
\begin{proof*}[Proof of Theorem  \ref{th:universal}]
By contradiction:
\begin{enumerate}
\item Assume there exists an infinite derivation $S \rightarrowtail^* \ldots$
\item Since $S \rightarrowtail^* T$, there exists a ground substitution
    $\theta$ satisfying Lemma~\ref{lemma:subs}.
\item Then
    $\mathtt{False} \leftarrowtail^* \theta.S \rightarrowtail^* \theta.T$
    by Lemmas~\ref{lemma:inf} and \ref{lemma:subs}.
\item Since $\theta.S$ is ground and $\theta.T \neq \mathtt{False}$,
    then $P$ is not ground-confluent. \label{case:non_conf}
\item Since $P$ is locally-confluent, $P$ is ground-locally-confluent.
\item \label{case:conf}
    Since $P$ is ground-locally-confluent and ground-terminating, $P$ is
    ground-confluent.
\item Contradiction between \ref{case:non_conf} and \ref{case:conf}. \quad
$\mathproofbox$
\end{enumerate}
\end{proof*}

Note that Theorem~\ref{th:universal} cannot be extended to the case
where $S \rightarrowtail^* \mathtt{False}$.
For example, consider the CHR program
\begin{align*}
    \mathtt{P}~x ~\simp~ \mathtt{False} ~~~~ ~~~~ ~~~~
    \mathtt{P}~x ~\simp~ x = \texttt{[}y\texttt{]},~\mathtt{P}~y
\end{align*}
This program is range restricted, ground-terminating, and locally confluent
since the only critical pair
$(\mathtt{False} \leftarrowtail \mathtt{P}~x \rightarrowtail
    x = \texttt{[}y\texttt{]},~\mathtt{P}~x)$ is join-able.
Although $(\mathtt{P}~x)$ is existentially terminating, i.e.
$(\mathtt{P}~x \rightarrowtail \mathtt{False})$, it is not universally
terminating because of the infinite derivation
$(\mathtt{P}~x \rightarrowtail x = \texttt{[}y\texttt{]}, \mathtt{P}~y
    \rightarrowtail x = \texttt{[}y\texttt{]}, y = \texttt{[}z\texttt{]},
        \mathtt{P}~z \rightarrowtail \ldots)$.
If we change the first rule $\mathtt{P}~x ~\simp~ \mathtt{True}$
(or any other non-$\mathtt{False}$ body), the program becomes
non-locally-confluent.

\subsection{Observable Confluence w.r.t. Existential Termination}

What remains is to establish confluence for terminating goals.
For notational convenience, we define $\mathsf{T}_\forall(S)$ and
$\mathsf{T}_\exists(S)$ to respectively hold if
state $S$ is universally or existentially terminating.
Clearly $\mathsf{T}_\forall$ is an observable invariant.
\begin{lemma}\label{lemma:uni_term}
If $P$ is locally-confluent, then $P$ is $\mathsf{T}_\forall$-confluent.
\end{lemma}
\vspace{-1em}
\begin{proof}
Define $\Sigma_\forall = \{S~|~S \in \Sigma \land \mathsf{T}_\forall(S)\}$,
then the abstract rewriting system
$\langle \Sigma_\forall, \rightarrowtail \rangle$ is
locally-confluent and terminating (by construction),
and is therefore confluent by a straightforward application of
Newman's Lemma~\cite{newman}.
\end{proof}
Alternatively, one can use the method from~\cite{duck07obs} to prove
$\mathsf{T}_\forall$-confluence.
However, this is overkill, as $P$ is already assumed to be locally confluent.

The condition $\mathsf{T}_\exists$ by itself is \emph{not} an
observable invariant,
since an existentially terminating state can be rewritten into a
universally non-terminating state.
However, if we define $\mathsf{T}'_\exists(S)$ to mean existential
termination to a non-false state, i.e.
$\mathsf{T}'_\exists(S)$ holds iff
there exists a derivation $S \rightarrowtail^* T \neq \mathtt{False}$
and $T \not\rightarrowtail$,
then we can state the following:
\begin{corollary}\label{cor:conf}
Let $P$ be a range-restricted, ground-terminating and locally-confluent
program, then $P$ is (1) $\mathsf{T}'_\exists$ is an observable invariant,
and (2) $\mathsf{T}'_\exists$-confluent.
\end{corollary}
\vspace{-1em}
\begin{proof}
By Theorem~\ref{th:universal},
$\mathsf{T}'_\exists = \mathsf{T}_\forall$, and therefore (1) holds.
By Lemma~\ref{lemma:uni_term}, $P$ is $\mathsf{T}_\forall$-confluent,
and therefore (2) holds.
\end{proof}

To elaborate further:
given a state $S$, suppose we execute $S$ and discover a
finite derivation $S \rightarrowtail^* T$, then $T$ is the only
possible answer for $S$.
\begin{corollary}[Uniqueness of Answers]\label{corr:unique}
Let $P$ be a range-restricted, ground-term\-inat\-ing and locally-confluent
program, if $S \rightarrowtail^* T \not\rightarrowtail$, then
for all $S \rightarrowtail^* U \not\rightarrowtail$ we have that
$T = U$.
\end{corollary}
\vspace{-1em}
\begin{proof}
If $\mathsf{T}'_\exists(S)$ then $T = U$ follows from 
Corollary~\ref{cor:conf}.
If $\neg \mathsf{T}'_\exists(S)$ then $T = U = \mathtt{False}$
since $\mathsf{T}_\exists(S)$ holds by assumption.
\end{proof}
Note that uniqueness of answers is not equivalent to confluence for
non-terminating programs.
For example, if
$(\ldots \leftarrowtail^* T \leftarrowtail^* S
    \rightarrowtail^* U \rightarrowtail^* \ldots)$
and if $T, U$ are non-joinable and universally non-terminating,
then $P$ is not confluent.
But $P$ may still produce unique answers for terminating goals.

\section{Practical Implications for Type Classes} \label{sec:haskell}

Type inference with type class constraints is an important application
of CHR.
Previously, strong conditions must be imposed in order to guarantee the
consistency, confluence and termination of type inference.
As will be explained in this section, the results from
Sections~\ref{sec:consistency} and~\ref{sec:confluence-terminating-goals}
allow for the relaxation of some of these conditions, which in turn,
allows for more programs to be safely accepted.

First, we summarize the standard translation scheme
from type classes to CHR, as well as the strong conditions required
for termination and confluence.
The remainder of this section discusses the relaxed conditions based
on our earlier results.

\subsection{From Type Classes to CHR}\label{sec:types_chr}

The basic syntax for a \texttt{class}-declaration is:
\begin{align}
\texttt{class}~D~\texttt{=>}~C~a_1~\ldots~a_n\texttt{|}~\mathit{fd}_1, \ldots,
    \mathit{fd}_n \tag{\textsc{Class}} \label{eq:class}
\end{align}
The declaration defines a new type-class $(C~a_1~\ldots~a_n)$ where
$a_i$ is a (type-variable) argument type.
Here, $D$ is a set of (super) type-class constraints for which
$C$ depends, and $\mathit{fd}_i$ is a \emph{functional dependency}
of the form $a_{i_1}, \ldots a_{i_k} \texttt{->} a_{i_0}$ where
$\{i_0, .., i_k\} \subseteq 1..n$.
Both $D$ and the $\mathit{fd}$ set may be empty and omitted.
The basic syntax for \texttt{instance}-declarations\footnote{
Both class and instance declarations also provide function interfaces
and implementations respectively.
However, these are not relevant to type inference, so we shall ignore
them here.
}
is:
\begin{align*}
    \texttt{instance}~D~\texttt{=>}~C~t_1~\ldots~t_n \tag{\textsc{Instance}}
    \label{eq:instance}
\end{align*}
Here $D$ is a set of type-class constraints for which the instance depends,
and $t_i$ are bound types.

For example, the following class declaration defines a $(\mathtt{Coll}~c~e)$
type-class constraint representing an abstract \emph{collection-type} $c$
with \emph{element-type} $e$:
\begin{align*}
\texttt{class Coll c e | c -> e} ~~~~ ~~~~ \texttt{instance Eq a => Coll [a] a}
\end{align*}
Here the class declaration states that the element type $e$ is
\emph{functionally dependent} on the collection type $c$, for more
formally: for all $a, b, c$, if $(\mathtt{Coll}~a~b)$ and
$(\mathtt{Coll}~a~c)$ then $b = c$.
The instance declaration states that $(\mathtt{Coll}~[a]~a)$ holds for any
type satisfying $(\mathtt{Eq}~a)$.

Both class and instance declarations can be understood as syntactic sugar
for collections of CHR rules~\cite{DBLP:journals/jfp/SulzmannDJS07}.
The basic translation schema is as follows:
For \texttt{class}-declarations of the form (\ref{eq:class}) we generate
the following rules:
\begin{align}
C~a_1~\ldots~a_n ~\prop~ & D \tag{\textsc{Class-Rule}} \label{eq:class_rule} \\
C~a_1~\ldots~a_n, C~b_1~\ldots~b_n ~\prop~ & a_{i_0} = b_{i_0}
    \tag{\textsc{FD-Rule}} \label{eq:fd_rule}
\end{align}
One (\ref{eq:fd_rule}) is generated for each $\mathit{fd}_i$ (of the
form $a_{i_1}, \ldots a_{i_k} \texttt{->} a_{i_0}$).
Here $b_j$ is $a_j$ if $j \in \{{i_1}, \ldots {i_k}\}$, otherwise
$b_j$ is a fresh variable.
Instance-declarations of the form (\ref{eq:instance}) generate the following
rules:
\begin{align}
C~t_1~\ldots~t_n ~\simp~ & D \tag{\textsc{Instance-Rule}}
    \label{eq:instance_rule} \\
C~b_1~\ldots~b_n ~\prop~ & b_{i_0} = t_{i_0} \tag{\textsc{Improvement-Rule}}
    \label{eq:improv_rule}
\end{align}
One (\ref{eq:improv_rule}) is generated per $\mathit{fd}_i$ provided
$t_j$ is not a variable.
Here $b_j$ is $t_j$ if $j \in \{{i_1}, \ldots {i_k}\}$, otherwise
$b_j$ is a fresh variable.
For example, the CHRs generated by the declarations for \texttt{Coll}
are:
\begin{align*}
\texttt{Coll}~c~e,~~\texttt{Coll}~c~d \prop e=d ~~~~ ~~~~ 
\texttt{Coll}~\texttt{[}c\texttt{]}~e \prop e=c  ~~~~ ~~~~
\texttt{Coll}~\texttt{[}a\texttt{]}~a \simp \texttt{Eq}~a
\end{align*}
It is possible to combine the last two rules
into a single rule 
$\texttt{Coll}~\texttt{[}a\texttt{]}~e \simp e=a, \texttt{Eq}~a$
as we have done in the introduction.

\subsection{Strong Conditions to guarantee Sound and Decidable Type Classes}

In order for type inference to be both \emph{sound} and \emph{decidable},
the resulting CHR rules must be consistent,
confluent and terminating.
If we allow for arbitrary class and instance declarations, this will not
always be the case.

Earlier work~\cite{DBLP:journals/jfp/SulzmannDJS07} identifies a
set of conditions that guarantee that the resulting CHR rules are both
terminating and confluent.
The CHR resulting from instance declarations must be terminating
and class declarations must satisfy the following two conditions:
\begin{itemize}[noitemsep,topsep=0pt,parsep=0pt,partopsep=0pt]
\item (\emph{Consistency Condition})
Consider a pair of instance declarations for a class \texttt{TC}:
\begin{align*}
\texttt{instance} ~D_1~ \texttt{=> TC}~ t_1~\ldots~t_n ~~~~ ~~~~
\texttt{instance}~ D_2~ \texttt{=> TC}~ s_1~\ldots~s_n
\end{align*}
Then, for each functional dependency
$\mathit{fd}_i = (a_{i_1}, ..., a_{i_k} ~\texttt{->}~ a_{i_0})$ for \texttt{TC},
the following condition must hold: for any substitution $\theta$ such that 
$\theta(t_{i_1}, ..., t_{i_k}) = \theta(s_{i_1}, ... , s_{i_k})$
we must have that $\theta(t_{i_0}) = \theta(s_{i_0})$.
\item (\emph{Coverage Condition})
Consider an instance declaration for class \texttt{TC}:
\begin{align}
\texttt{instance} ~D~ \texttt{=> TC}~ t_1~\ldots~t_n    \label{eq:inst_decl}
\end{align}
Then, for each functional dependency
$\mathit{fd}_i = (a_{i_1}, ..., a_{i_k} ~\texttt{->}~ a_{i_0})$ for \texttt{TC},
we require that $\vars(t_{i_0}) \subseteq \vars(t_{i_1},\ldots,t_{i_k})$.
\end{itemize}

\subsection{Relaxed Conditions to guarantee Soundness for Terminating Goals}

Many practical programs violate the Coverage Condition.
Recall the program
\begin{verbatim}
class F a b | a -> b   instance F Int Bool   instance F a b => F [a] [b]
\end{verbatim}
which violates the Coverage Condition
because $\vars(\texttt{[b]}) \not\subseteq \vars(\texttt{[a]})$.

We cannot naively drop the Coverage Condition; but we may impose
the following \emph{Weak Coverage Condition}.
\begin{itemize}[noitemsep,topsep=0pt,parsep=0pt,partopsep=0pt]
\item (\emph{Weak Coverage Condition})
For the instance declaration (\ref{eq:inst_decl}) and
each functional dependency
$\mathit{fd}_i = (a_{i_1}, \ldots, a_{i_k} ~\texttt{->}~ a_{i_0})$,
then
$\vars(t_{i_0}) \subseteq \mathit{closure}(D, \mathit{vs})$
where
\bda{lcl}
\mathit{closure}(D, \mathit{vs}) & = & \bigcup_{i=1}^{\infty} \mathit{covered}^i(D,vs)
\\
\mathit{covered}^1(D,vs) & = & \bigcup_{\ba{l} \texttt{TC}~t_1\ldots t_n \in D \\  
                             \makebox[0pt][l]{$\texttt{TC}~a_1\ldots a_n \mid a_{i_1}, ..., a_{i_k} ~\texttt{->}~ a_{i_0}$}   \ea}
         \{ \tv(t_{i_0}) \mid \tv(t_{i_1},\ldots,t_{i_k}) \subseteq vs \}
\\
\mathit{covered}^{i+1}(D,vs) & = & \mathit{covered}^1(D,\mathit{covered}^i(D,vs))
\eda
\end{itemize}
Like the Coverage Condition, the
Weak coverage is sufficient to establish \emph{local confluence} of the
resulting CHR rules
in combination with the
Consistency Condition~\cite{DBLP:journals/jfp/SulzmannDJS07}.
However, unlike the Coverage Condition, Weak Coverage 
is not sufficient to establish \emph{termination}.
Recall the infinite derivation from the introduction
\bda{l}
\mathtt{F}~\texttt{[}a\texttt{]}~a ~\rightarrowtail~
(\mathtt{F}~\texttt{[}b\texttt{]}~b,~a = \texttt{[}b\texttt{]})
    ~\rightarrowtail~
(\mathtt{F}~\texttt{[}c\texttt{]}~c,~a = \texttt{[}b\texttt{]},~
    b = \texttt{[}c\texttt{]}) ~\rightarrowtail~
\ldots
\eda
Fortunately, such devious goals usually do not show up for realistic programs.

We can summarize the \emph{relaxed conditions} as follows.
Given a set $C$ of class and instance declarations, we derive the
corresponding CHR program $P$ from $C$ using the translation from
Section~\ref{sec:types_chr}.
The relaxed conditions are essentially the same as that used by our
CHR theoretical results, namely
\begin{itemize}[noitemsep,topsep=0pt,parsep=0pt,partopsep=0pt]
    \item (\emph{Range restrictedness}): $P$ must be range restricted;
    \item (\emph{Local Confluence}): $P$ must be locally-confluent; and
    \item (\emph{Ground Termination}): $P$ must be ground-terminating.
\end{itemize}

Range restrictedness of $P$
can be established via simple syntactic checks.
For example, if all given instance declarations of the form
$(\texttt{instance}~\mathit{Ctx} \Rightarrow H)$ satisfy the constraint
$\vars(\mathit{Cxt}) \subseteq \vars(H)$, then the resulting
$P$ will be range-restricted~\cite{DBLP:journals/jfp/SulzmannDJS07}.
Local confluence follows directly from the Weak Coverage Condition and
the Consistency Condition~\cite{DBLP:journals/jfp/SulzmannDJS07}.

To prove Ground Termination we can rely on the
existing state-of-the-art work on
termination for CHR programs, such
as~\cite{thom_termination} and~\cite{Pilozzi08iclp}.
For example, we can prove that the rule
$(\mathtt{F}~\texttt{[}a\texttt{]}~b \simp b=\texttt{[}c\texttt{]},~
\mathtt{F}~a~c)$ is ground terminating by defining
$\mathit{rank}(\texttt{[}x\texttt{]}) = 1 + \mathit{rank}(x)$.  Each
rule application to a ground state decreases the rank, so any
corresponding derivation must eventually terminate.

An alternative method for proving ground termination in our context is
the notion of CLP projection as described in~\cite{HLH11ppdp}.
Formally, the {\em projection} of a simplification rule $(h_1,\dots,
h_n \simp B)$ is the set of Horn clauses $\{h_i \leftarrow B \mid i
\in 1, \dots n\}$.  The projection of a CHR program is the union of
the projections of its simplifications. %
If the projection of a set $P$ of mono-headed simplifications is
terminating then so is $P$~\cite{HLH11ppdp}.  Since purely built-in
propagation rules either do not apply or fail on ground states, there
exists a direct correspondence between the ground termination if $P$
and its projection.  We can therefore use state-of-the-art CLP
termination analysis tools to verify ground-termination of the CHR
type inference programs. %
For instance, we used the AProVE analyzer~\cite{GSTijcar06} to
automatically prove ground-termination of all the programs given as
examples in the present paper.

\subsection{Correctness of the \texttt{UndecidableInstances} Flag}

Assuming the relaxed conditions are satisfied, we can verify the
correctness of type inference in GHC 
under the \texttt{UndecidableInstances} flag.
We can formalize the behavior of this flag as follows:
given a depth bound $B$ and a goal $S$, we choose a bounded derivation
$S \rightarrowtail S_1 \rightarrowtail \dots \rightarrowtail S_b$ for $S$
such that either:
\begin{itemize}[noitemsep,topsep=0pt,parsep=0pt,partopsep=0pt]
    \item[-] (\emph{Final State}) $S_b \not\rightarrowtail$, $b \leq B$, then
        the answer is $S_b$; or
    \item[-] (\emph{Unknown}) $S_b \rightarrowtail \ldots$, $b = B$, then
        the answer is \emph{unknown}.
\end{itemize}
An answer of \emph{unknown} is reported to the user in the form of a compiler
error.
Otherwise, by Corollary~\ref{corr:unique} we know that $S_b$ is the one
unique answer for any finite derivation of $S$.

\section{Conclusion and Related Work} \label{sec:conc}

The idea that confluent programs are consistent can be traced back
to early CHR confluence results~\cite{Abdennadher99confluence},
but the general proof is more recent~\cite{HLH11ppdp}. %
In comparison with these earlier works, the main result of 
Section~\ref{sec:consistency} requires a weaker form of confluence (i.e.\
ground confluence) in combination with the additional condition that
CHR are range-restricted. %
In the context of types, consistency is an important condition
to guarantee type safety (``well-typed programs will not go wrong'').
Hence, the result of Section~\ref{sec:consistency} 
provides some general consistency criteria to ensure that
type class programs are safe.

Establishing confluence in the presence of non-termination is a
notoriously difficult problem~\cite{Haemmerle12iclp}.  Our results in
Section~\ref{sec:confluence-terminating-goals} advance the state of
the art in this area by showing that existentially-terminating goals
(to non-\texttt{False} states)
are confluent for range-restricted, ground-terminating and locally
confluent programs.  These results have an important practical
applications in the type inference setting for type classes.

In our current formulation, the ground termination assumption
trivially rules out super classes, i.e.\ CHR rules
which propagate user constraints.
Range-restrictedness rules out instance declarations such as
(\verb+instance (F a c, F c b) => F [a] [b]+)
since variable \texttt{c} does not appear in \texttt{F [a] [b]}.
We believe that it is possible to relax both restrictions.
This is something we plan to investigate in future work.

In another direction, we intend to investigate
to what extent our results are transferable to
type functions~\cite{DBLP:conf/icfp/SchrijversJCS08},
a concept related to type classes with functional dependencies.

From the point of view of general CHR confluence state of art, we plan
generalizing consistency of ground-confluent but non range-restricted
program by using CLP projection~\cite{HLH11ppdp}.  It seems also worthwhile to prove
ground-confluence of non ground-terminating programs using
diagrammatic techniques~\cite{Haemmerle12iclp}.

\section*{Acknowledgments}

We thank the reviewers for their comments.

\bibliography{main}

\end{document}